\documentstyle[aps,epsfig,subfigure,cite]{revtex}
\begin{document}
\draft
\title{Properties of Layer-by-layer Vector Stochastic Models of Force
       Fluctuations in Granular Materials}
\author{M. L. Nguyen and S. N. Coppersmith}
\address{The James Franck Institute and Department of Physics\\
         The University of Chicago, 5640 S. Ellis Ave., Chicago, IL 60637}
\date{\today}
\maketitle
\begin{abstract}
We attempt to describe the stress distributions of granular
packings using lattice-based layer-by-layer stochastic models
that satisfy the constraints of force and torque balance and non-tensile
forces at each site.  The inherent asymmetry in the
layer-by-layer approach appears to lead to an asymmetric force
distribution, in disagreement with both experiments and general
symmetry considerations.  The vertical force component 
probability distribution is robust and in agreement with predictions
of the scalar $q$ model~\cite{Liu:q:1995, Coppersmith:q:1996} while
the distribution of horizontal force components is qualitatively
different and depends on the details of implementation.
\end{abstract}
\pacs{}

\widetext
\section{Introduction}
\label{sec:Intro}

Gaining an understanding of the inhomogeneous stress distribution in a
granular packing is important both because of the insight
it may yield into failure mechanisms and as a first step towards
elucidating the dynamics of such systems~\cite{Dantu:1967, Drescher:1972,
Travers:1987, Edwards:1996a, Mounfield:1996, Edwards:1996b,
Wittmer:1996, Wittmer:1997, Claudin:1998, Hemmingsson:1996,
Hemmingsson:1997}. Although qualitative aspects of the
inhomogeneities have been known for some time~\cite{Ammi:1987b,
Liu:q:1995, Baxter:1997, Dantu:1957, Dantu:1967, Wakabayashi:1959,
Travers:1988, Howell:1997}, quantitative experiments have been
performed only recently for cylindrical packings of glass
beads~\cite{Mueth:1998}, 2D arrangements of optical
fibers~\cite{Baxter:1997}, and shear cells of glass
spheres~\cite{Miller:1996}. These experiments, together with  
numerical simulations~\cite{Radjai:1996, Thornton:1998, Clelland:unpub},
have yielded new insight into the nature of stress inhomogeneities.
Various properties of the stress distributions are consistently
observed both in two-~\cite{Radjai:1996, Baxter:1997, Clelland:unpub} and
three-dimensional~\cite{Liu:q:1995, Mueth:1998, Thornton:1998}
systems.  One such feature is that $P(f)$, the
probability of observing a force $f$, decays exponentially with $f$
for forces much larger than the mean~\cite{Liu:q:1995, Mueth:1998,
Radjai:1996}.

A statistical model for a single component of stress in a packing
based on a layer-by-layer approach appears to capture some features
of the observed fluctuations~\cite{Liu:q:1995, Coppersmith:q:1996}.
In this model, the disorder in the system leading to the creation of
force chains, whether from the inhomogeneities in the packing itself,
variations in the size of the particles, or differences in material
properties, is encapsulated in a set of random variables labeled
$q_{ij}$ which determine the fraction of the stress component is
transferred from an element $i$ of the packing to a neighboring
element $j$.  In \cite{Liu:q:1995} and \cite{Coppersmith:q:1996},
these fractions are chosen randomly from probability distributions
consistent with the constraint of force balance with the assumption
that the $q$'s at different sites are uncorrelated.  Correlated $q$
models have also been investigated~\cite{Claudin:1997, Nicodemi:1998}.
The layer-by-layer structure enables analytic progress on
characterizing the force distributions.  The predicted exponential
decay in the probability distribution for large values of the stress
component  is in qualitative agreement with experimental and
simulation results. 

The $q$ model is scalar: it ignores the contributions that balancing the
remaining stress components and torque may have in determining the
weight redistribution and yields no information on their
probability distributions.  Investigation of correlations between
fluctuations of shear and of compression effects, processes which are
important in understanding failure, are not possible with this model.
Moreover, though the $q$ model successfully describes the stress
fluctuations in simple geometries where the large-scale stresses are
spatially constant, if applied to situations where stress varies over
long scales, it predicts that these variations should obey a diffusion
equation, in disagreement with experiment~\cite{Wittmer:1997,
Edwards:1996a}.  These shortcomings have led several groups to examine
stress fluctuations in models which incorporate vector
forces~\cite{Eloy:1997, Socolar:1998, Pitman:1998, Kenkre:1998a,
Scott:1998b}. 

The generalization of the scalar $q$ model to vector forces is an
important step in the understanding of fluctuations about
locally-averaged quantities calculated using continuum theories.
Claudin \textit{et al.}~\cite{Claudin:1998} have investigated the
connection between the lattice-based $q$ model and the ``light-cone''
continuum equations of Wittmer \textit{et al.}~\cite{Wittmer:1997} by
examining continuum equations with randomness.  This connection is
not trivial.  In addition to subtleties encountered when one takes the
continuum limit of stochastic models, Claudin \textit{et al.} expose
an important complication that arises when significant randomness is
introduced into their continuum equations--- the occurrence
of tensile forces.  They interpret this result, quite reasonably,
as indicating that the granular material must rearrange.  However,
the materials should eventually reach a state at which the load should
be supported and all the forces non-tensile.  It is unclear how to
describe this state using their approach.  Generalization of the
lattice-based models provides another means of approaching the issue.

Describing the system using random variables subject to the constraints
of force balance at each site and no tensile forces provides a
mechanism by which more realistic vector models may be based.  Vector
models in this spirit have been proposed by Eloy and
Cl\'{e}ment~\cite{Eloy:1997} and Socolar~\cite{Socolar:1998}.
Common to the proposals is the propagation of forces downward in the
packing in a layer-by-layer fashion starting at a load applied at its
top. 

In this paper we present our attempts to construct a layer-by-layer
vector force model.  We find that serious fundamental problems arise
from attempts to describe vector force fluctuations using simple
generalizations of the $q$ model.  In particular, we find that it is
very difficult to construct a stochastic model that leads to isotropic
force fluctuations and satisfies the constraints of force and torque
balance with non-tensile forces.  By symmetry, isotropy of
these fluctuations is expected when a system is both prepared and 
compressed isotropically; moreover, Mueth \textit{et 
al.}~\cite{Mueth:1998} demonstrate experimentally that subjecting a
granular system to a uniform isotropic pressure leads to a stress
fluctuation distribution which appears to be isotropic.  The difficulty
arises because of the inherent asymmetry in the formulation of a
layer-by-layer vector model.  At the individual site level, we see the
possibility of the creation of large-magnitude horizontal output
forces that are independent of the input forces and torque.  The
layer-by-layer structure does not allow for an intrinsic mechanism of
removal, leading the most natural formulations of the model to have
horizontal forces that appear to grow without bound as the system
depth is increased.  Incorporating cutoffs on the force magnitudes
appears necessary to achieve any set of reasonable force
redistributions.  We find the probability distributions of vertical 
forces are insensitive to the cutoffs while the probability
distribution of horizontal forces are dependent on their form.
Thus, these cutoffs do not appear to present a solution to the underlying
pathology of the model.  

The paper is organized as follows.  Section~\ref{sec:Model} defines the
model we investigate.  We consider the redistribution of forces at
a single site and discuss the difficulties that may be seen even at
this level.  Section~\ref{sec:Methods} describes the process of
redistributing forces through a lattice including our strategies for
limiting the magnitudes of the forces generated by the redistribution
algorithm.  Section~\ref{sec:Results} reports the results of the
statistics of the forces obtained for forces obtained for various
choices of the coefficient of friction $\mu$ and force cutoffs,
illustrating the inherent asymmetry and cutoff dependence.
Section~\ref{sec:Discussion} compares the results to experiment.
Appendix~\ref{app:PFB} describes methods used to increase the efficiency
of generating force redistributions.  Appendix~\ref{app:Four}
discusses exactly solvable four-site lattice configurations to gain
insight into strategies that could be used to construct isotropic
models.

\section{Model}
\label{sec:Model}

\subsection{Force Balance at a Site}
As in the scalar version of the $q$ model~\cite{Liu:q:1995,
Coppersmith:q:1996}, we assume that the essential features of the  
disorder in a granular packing can be described using random
variables.  The choice of these variables is constrained by the
requirements of satisfying force and torque balance as well as the
requirement that the forces be non-tensile.  We describe below the
specific representation of random variables that we have employed.  
For the model to be considered useful, the probability distributions
of vertical and horizontal force components should not be 
sensitive to the details of these choices.

\begin{figure}
\centering
\epsfig{file=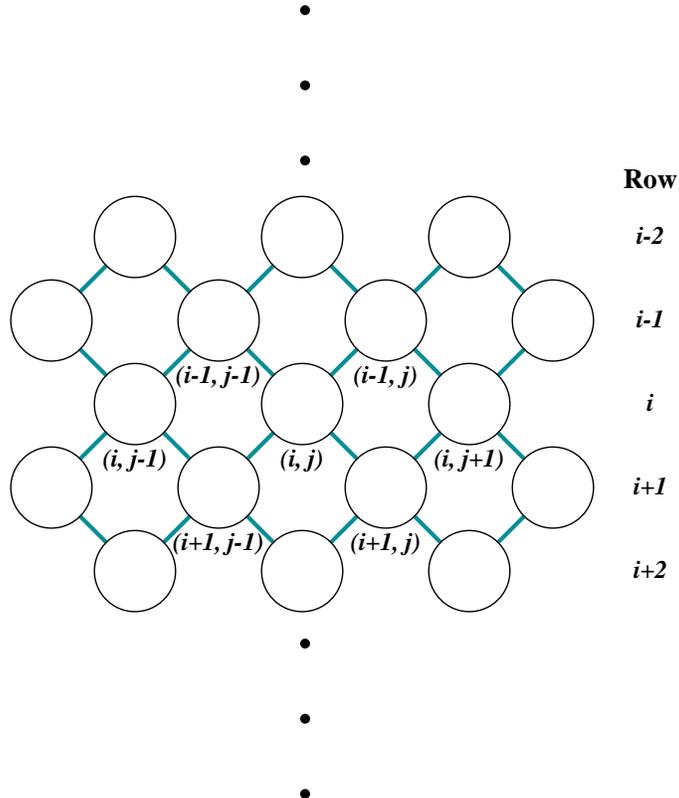, width=3.5in}
\caption{A triangular lattice is used to approximate a 2-D granular
packing in order to assign depth in site and neighboring sites.  Lines
connecting sites indicate that a transfer of force occurs between them.
Sites on the same layer are not connected.  Site $(i,j)$ is labeled
along with its neighboring sites.}
\label{fig:lattice}
\end{figure}

In our implementation of a layer-by-layer vector model we assume the
topology of the packing is that of a modified regular two-dimensional
triangular lattice of discs, as shown in figure \ref{fig:lattice}.  
Forces are introduced at the top of the lattice and are propagated
downwards with ``input'' forces at a site arising from the two
neighbors in the layer above and the resulting ``output'' forces being
passed on to the two neighbors below.  Sites on the same layer do not
transfer force between one another.  In an arbitrary $N$ row by $M$
column lattice, the $j$th site in the $i$th layer transmits its
leftward output to the site $j-1$ ($j$) on the $i+1$ layer and its
rightward to $j$ ($j+1$) for odd (even) $i$. 

\begin{figure}
\centering
\epsfig{file=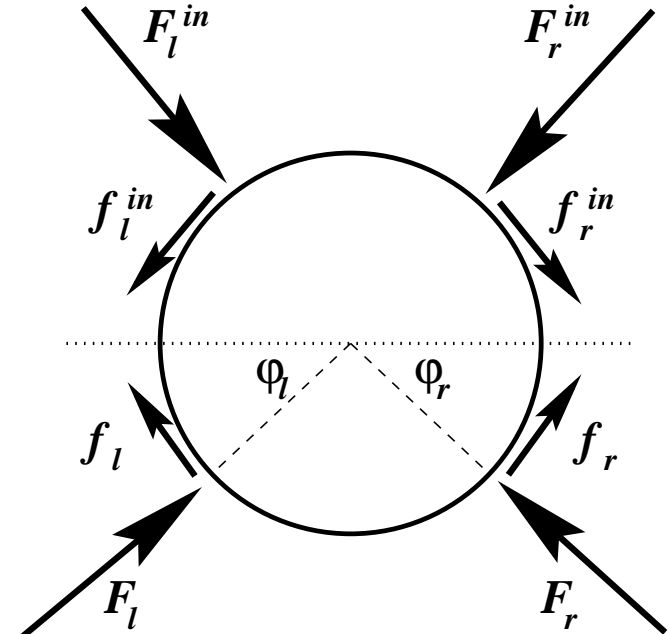, width=3.5in}
\caption{Schematic of forces at a site for the vector model.  Input
forces are from the neighboring sites above and output forces are from
the sites below.  The frictional forces have magnitude equal to 
$f_{l,r} = \mu \eta_{l,r} F_{l,r}$, where $F_{l,r}$ is the normal
force, and are shown in the $+\eta$ direction.  The angles
$\varphi_{l,r}$ indicate the contact angle between the site and the
output neighbors.  The parameters $\varphi_{l,r}$ and $\eta_{l,r}$
are chosen randomly, consistent with the constraints of force and torque
balance, non-tensile forces, and $|\eta_{l,r}| \leq 1$.}
\label{fig:site}
\end{figure}

We have chosen to parameterize the redistribution of forces at a site
by randomly chosen contact angles and effective friction coefficients.
Figure \ref{fig:site} shows a schematic of the forces acting on a
site.  The contact angles $\varphi_{l,r}$ determine the direction of
the output normal forces $F_{l,r}$ to the left and right neighbors,
respectively, and are chosen from the interval $(0, \frac{\pi}{2})$.
The effective friction coefficients $\eta_{l,r}$ determine the
direction and magnitude of the output tangential forces and are chosen
from the interval $[-1,1]$.  The magnitude of the tangential force
$f_{d}$ ($d=l,r$) varies from $[0,|\mu \eta_d F_d|]$ with $\mu$ being
the coefficient of static friction.  Positive values of $\eta_d$
have tangential force $f_d$ contributing to the balancing of the
vertical force component at a site in conjunction with its paired
normal $F_d$, while negative values have the tangential force in
opposition to the paired normal.  The tangential forces contribute to
the torque $\Gamma R$ at a site and are assumed to occur at the same
distance $R$ from its center of mass.  The constraints imposed by
force and torque balance and the non-tensile force requirement will
place further restrictions on the allowed range of the random
variables.

In any stationary packing, the individual sites must satisfy force and
torque balance: 
\begin{mathletters}
\label{eq:balance}
\begin{eqnarray}
F_{x}^{in} & = & - F_{l} (\cos \varphi_{l} - \mu \eta_{l} \sin \varphi_{l})
                 + F_{r} (\cos \varphi_{r} - \mu \eta_{r} \sin \varphi_{r}),
\\
F_{y}^{in} & = & F_{l} (\sin \varphi_{l} - \mu \eta_{l} \sin \varphi_{l})
                 + F_{l} (\sin \varphi_{r} - \mu \eta_{r} \sin \varphi_{r}),
\\
\Gamma^{in} & = & \mu (\eta_{l} F_{l} - \eta_{r} F_{r}),
\label{eq:torque_balance}
\end{eqnarray}
\end{mathletters}
\noindent where the total input force components and torque are given by
the sum of the normal and frictional forces from neighboring sites
in the layer above:
\begin{mathletters}
\begin{eqnarray}
F_{x}^{in} & = & \left( \mathbf{F_{l}^{in}} \right)_{x} + 
	         \left( \mathbf{F_{r}^{in}} \right)_{x}, \\
F_{y}^{in} & = & \left( \mathbf{F_{l}^{in}} \right)_{y} + 
	         \left( \mathbf{F_{r}^{in}} \right)_{y}, \\
\Gamma^{in} & = & \Gamma_{l}^{in} + \Gamma_{r}^{in}.
\end{eqnarray}
\end{mathletters}
Only the total input force and torque enter into
eqs. (\ref{eq:balance}) as these are fixed values arising from the
propagation of forces in the previous layer.  Because of torque 
balance, eq. (\ref{eq:torque_balance}), the $\eta$'s are not
independent--- $\eta_{l}$ may be written in terms of $\eta_{r}$ (or
vice versa).  The number of independently chosen random
variables is reduced to three. 

Solving for the output normal forces yields:
\begin{mathletters}
\label{eq:forces}
\begin{eqnarray}
F_l & = \frac{1}{\Psi} &
      \left[ F_x^{in} \left(\sin \varphi_{r} +
                           \mu \eta_{r} \left(\cos \varphi_{l} +
                                              \cos \varphi_{r}\right)
                     \right)
            - F_y^{in} \left(\cos \varphi_{r} +
                             \mu \eta_{r} \left(\sin \varphi_{l} -
                                                \sin \varphi_{r}\right)
                       \right) \right. \nonumber \\
& & \ \ \left.+ \Gamma^{in} \left(\cos \left(\varphi_{l} +
                                            \varphi_{r}\right) -
                                 \mu \eta_{r} \sin \left(\varphi_{l} +
                                                         \varphi_{r}\right)
                           \right)
       \right], \\
F_r & = \frac{1}{\Psi} &
          \left[-F_x^{in} \sin \varphi_{l} - F_y^{in} \cos \varphi_{l}
                + \Gamma^{in}
          \right],
\end{eqnarray}
\end{mathletters}
where
\begin{equation}
\Psi = -\mu \eta_{r} \left[1 + \cos \left(\varphi_{l}
                                              + \varphi_{r}\right)\right]
           -\sin \left(\varphi_{l} + \varphi_{r}\right) .
\label{eq:psi}
\end{equation}
\noindent The asymmetry between the forms of $F_{l}$ and $F_{r}$ is due 
to the choice of designating $\eta_{r}$ as a random statistical variable.

\begin{figure}
\centering
\mbox{
\subfigure[]{
\epsfig{file=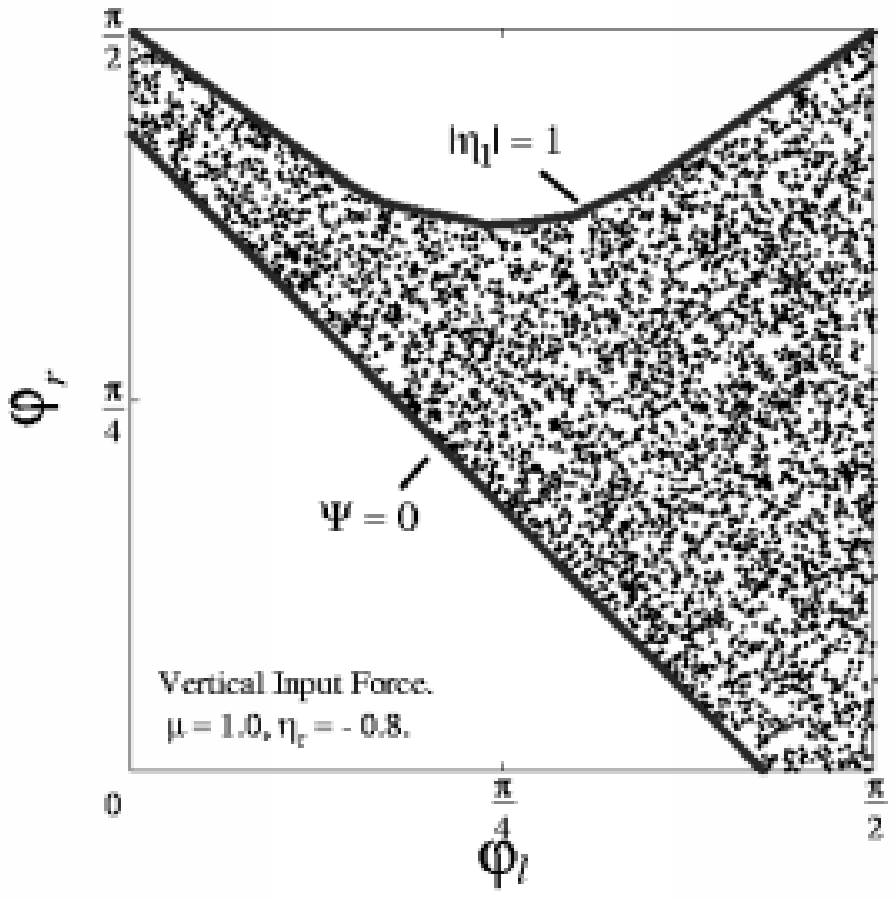, width=3.375in}}
\subfigure[]{
\epsfig{file=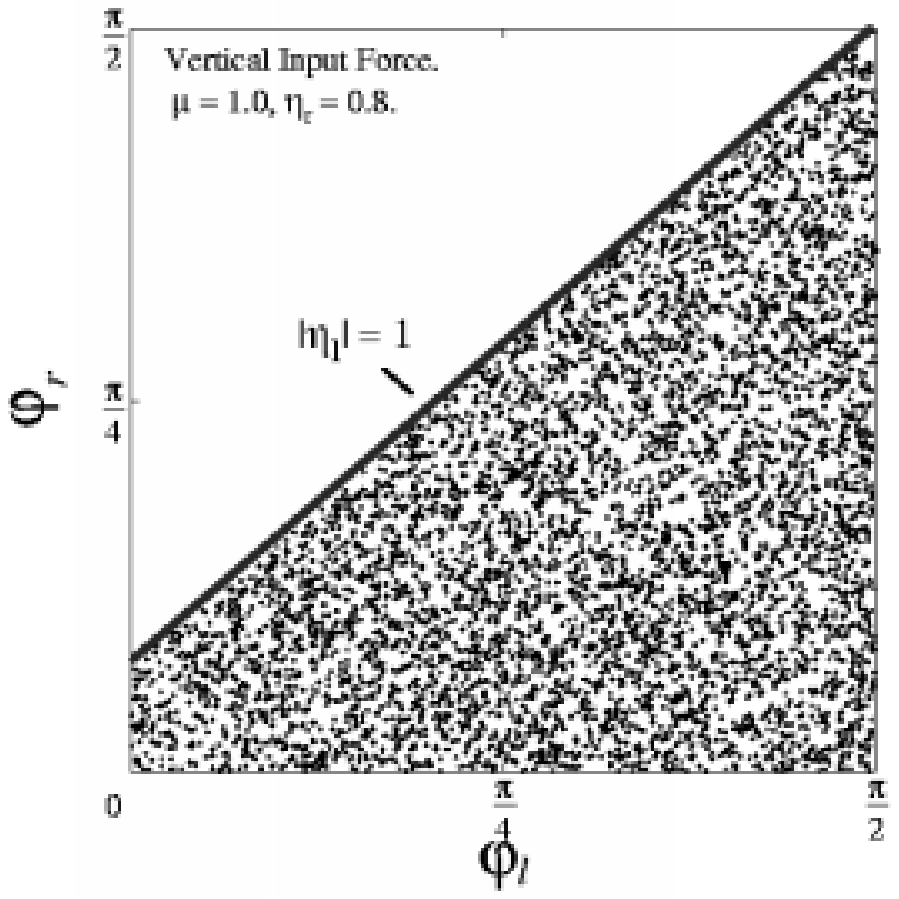, width=3.375in}}
}
\caption{Slices of configuration space of contact angles
$\varphi_{l,r}$ and effective friction coefficient $\eta_r$.  
The maximum coefficient of static friction $\mu$ has been set to 1 to
exaggerate the features.  A sample of 10,000 points were randomly
chosen with uniform probability in the interval $[0, \frac{\pi}{2}]
\times [0,\frac{\pi}{2}]$.  Points that satisfied the non-tensile and
frictional constraints, eqs. (\ref{eq:constraints}), are marked with a
$+$.  Boundaries of the valid region are labeled.  The $\Psi=0$
boundary where large horizontal forces are generated is found for
$\eta_r \leq 0$ but not for $\eta_r > 0$.}  
\label{fig:configuration}
\end{figure}

The force redistribution is considered valid if for the values of
$\varphi_{l,r}$ and $\eta_{r}$ chosen, 
\begin{mathletters}
\label{eq:constraints}
\begin{eqnarray}
F_{l,r} \geq 0 & & \ \ \mathrm{(Non\!\!-\!\!tensile\ constraint)},
\label{eq:non-tensile}\\
|\eta_{l}| \leq 1 & & \ \ \mathrm{(Newtonian\ friction)}.
\label{eq:friction}
\end{eqnarray}
\end{mathletters}
Figure \ref{fig:configuration} shows the valid region of contact
angle-effective friction coefficient configuration space for a vertical
input force with positive and negative values of $\eta_r$.  The
formalism for choosing $\eta_{l}$ as a random variable instead of
$\eta_{r}$ is similar.

\subsection{Difficulties Arising at a Site}

Even at the single-site level, we may encounter difficulties in the
redistribution and propagation of forces.  One possibility is that the
input forces and torque may be such that no valid redistribution
can occur.  Although the force and torque balance equations always 
yield solutions for the inputs in any randomly chosen
configuration of angles and friction, the additional non-tensile and 
frictional constraints may severely limit the number that may be realized.  
Typically, we see that the configuration space is significantly reduced 
if the net input force is largely horizontal or if the input torque is 
large relative to the input normal and frictional forces.  

A more serious difficulty arises because large magnitude output forces
can be generated irrespective of the magnitude of the input force.  To
illustrate this problem, we write the force balance at a site, given
by eqs. (\ref{eq:balance}), schematically as
\begin{equation}
\mathbf{F}^{in} = \mathbf{M}(\varphi_{l}, \varphi_{r}, \eta_{r})
                  \mathbf{F}_{normal}^{out},
\label{eq:fb_scheme}
\end{equation}
which has the solution
\begin{equation}
\mathbf{F}_{normal}^{out} = \mathbf{M}^{-1} \mathbf{F}^{in}.
\end{equation}
The factor $\Psi$ from eq. (\ref{eq:psi}) is the determinant 
of the matrix $\mathbf{M}$:
\begin{equation}
\Psi \equiv \det \mathbf{M}.
\end{equation}
As $F_{l,r} \propto \Psi^{-1}$, large magnitude
normal forces will be generated whenever $\Psi$ approaches zero.
$\Psi$ depends only on the choice of random variables and is independent 
of the input force.
 
\begin{figure}
\centering
\epsfig{file=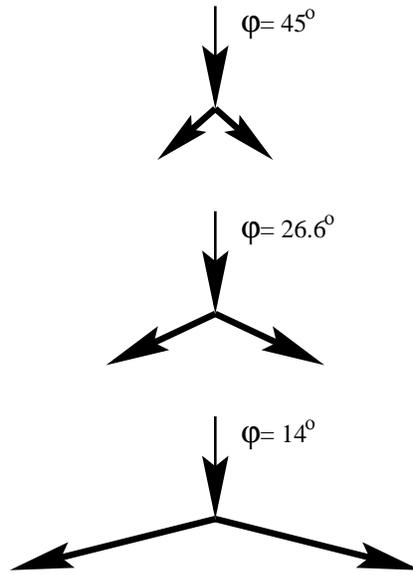, height=3in}
\caption{Even for a site with a purely vertical input force in the
absence of friction, the output forces must increase significantly in
magnitude as both output angles approach the horizontal.  Because the
horizontal force components in the outputs are in opposition to each
other, they may grow without bound while the vertical components are
limited in magnitude by the input.  We see this growth from a small
magnitude in (a) to a large magnitude in (c) for the same vertical
input force.  Values of $\Psi$ from eq. (\ref{eq:psi}) for each 
case are $\Psi_{45^{\circ}}=-1$, $\Psi_{26.6^{\circ}}=-0.8$, and
$\Psi_{14^{\circ}}=-0.47$.} 
\label{fig:large_mag}
\end{figure}

Exactly at the $\Psi = 0$ boundary, the output forces sum vectorially
to zero--- no force balance is possible.  This boundary occurs at
values of $\varphi_{l,r}$ and $\eta_{r} \leq 0$ satisfying
\begin{equation}
\cos (\varphi_{l} + \varphi_{r}) = \frac{1 - (\mu \eta_{r})^{2}}
                                        {1 + (\mu \eta_{r})^{2}}.
\label{eq:boundary}
\end{equation}
The boundary can be clearly seen in figure~\ref{fig:configuration}(a).
Although the boundary itself does not yield solutions, regions of
configuration space exist near it that do yield valid solutions for
which $\left| \Psi \right|$ is very small.  In these 
regions the output forces almost cancel, requiring large
magnitude normal forces in order to satisfy any equilibrium
condition for non-zero input values.  Because the input force is
designated as originating from the neighboring sites from the layer
above, an asymmetry exists between vertical and horizontal components.
The vertical components of the output forces are bounded in magnitude
by the fixed vertical component of the input while the horizontal
components of the leftward and rightward outputs acting in opposition
to each other are essentially unbounded.  This effect can easily be
seen in the non-frictional case for a purely vertical input force: as
shown in figure~\ref{fig:large_mag}, as the angle of both output
forces approach the horizontal, the magnitude of those forces must
increase so that their vertical components will balance
the input.  The frictional case is similar, though the addition of
tangential forces complicates the picture slightly.

The vector model proposed by Eloy and Cl\'{e}ment~\cite{Eloy:1997}
also suffers from this pathology although the mechanism may not be
as obvious due to their choice of parameterization.  Socolar's
model~\cite{Socolar:1998} requires the forces to lie in a $45^{\circ}$
cone about the vertical and therefore only considers non-negative
values of friction (as interpreted by our model).  As a result, it does
not generate these large forces but at the cost of imposing severe limits
on the magnitude of horizontal components.  We discuss this issue in
more detail below.

\section{Methods}
\label{sec:Methods}

\subsection{Force Redistribution through a Lattice}
Our algorithm for generating large lattices redistributes
the forces at individual sites starting from a normally distributed
vertical load applied on the top layer of sites and proceeding
downward into the lattice layer by layer.  All sites on a layer are
processed before proceeding to the next layer down.  Contact angles
$\varphi_{l,r}$ and an effective friction coefficient, either
$\eta_{l}$ or $\eta_{r}$, at a site are randomly chosen within
their respective intervals using a uniform deviate and a test is made
to determine whether the resulting force redistribution satisfies the
necessary non-tensile and frictional constraints.  If the constraints
are not met, new sets of random variables for the site are chosen
until all requirements are satisfied.  

Failure in lattice generation occurs when the input forces and torque
at a site cannot be be redistributed within a reasonable sampling of
the contact angles-effective friction coefficient configuration
space.  Reasonable has been defined as a sampling of 25,000
uniformly distributed points in the space.  A larger sampling size did
not increase the lattice yield significantly.  Furthermore, some input
force configurations do not have any valid redistributions.  In our
simulation, if no valid redistribution for a site is found, lattice
generation is terminated and restarted with another random number
seed.  While encountered mainly when friction has been applied, we
find that non-frictional cases may also suffer from this behavior.
The rate of failure increases with lattice size and coefficient of
static friction $\mu$.  However, techniques described in
appendix~\ref{app:PFB} can be used to increase the yield.

\subsection{Implementation of Limits on Force Magnitude}

As discussed above, large horizontal forces can be generated at a
single site.  However, we consider the possibility that the lattice
structure may be self-limiting so that the effect of large magnitude
horizontal forces will be restricted to small neighborhoods
surrounding the generating site.  One possible mechanism would be
the cancellation occurring at a site with left and right input
neighbors both contributing this type of force.  However, as shown
below, we find that the generation of a cancellation pair is unlikely
enough that these large forces build up and propagate in our
lattices.

Having failed to identify an intrinsic means to limit the
production and transmission of large magnitude horizontal forces,
we impose various cutoff schemes to attempt to generate sets of
realistic force  distributions.  The choice of one
form of implementation over another is somewhat arbitrary.
Consequently, several cutoff schemes have been implemented
to observe the influence they exert on the resulting force
distributions. 

The first cutoff scheme we implement is the simplest: the magnitude
of each normal force at a site may not exceed a specified value.
The second scheme is to limit the allowed angles to prevent
exploration of the $\Psi=0$ boundary.  This method is similar to the
Eloy and Cl\'{e}ment model~\cite{Eloy:1997} which has fixed angles.
By restricting the range of available angles, this type of limit
serves to reduce the volume of configuration space available for
redistribution and can exclude regions that form large 
magnitude horizontal forces.  Our third scheme is a soft cutoff scheme
based on the assumption of contact energies between sites following a
Boltzmann-like distribution.  This choice is clearly arbitrary as the
system is not thermal.  For simplicity, the lattice is modeled as a
layer of spheres whose centers lie on the same plane (contact topology
is therefore equivalent to discs) and elastic 
theory~\cite{Landau:Elastic} is used to calculate the energy $U$
within the contacts between these spheres:
\begin{equation}
\label{eq:contact_U}
U = A(\sigma, E, R) F^{\frac{5}{3}},
\end{equation}
where $F$ is the normal force between the sites in contact and $A$ is a
function of the material properties (Poisson's ratio $\sigma$ and Young's 
modulus $E$) and radius $R$ of the spheres.  We assume that the probability 
that a contact has an energy $U$ follows an exponential distribution 
\begin{equation} 
P(U) = \frac{1}{U_{0}} e^{-U/U_{0}},
\label{eq:soft}
\end{equation}
where $U_{0}$ is is an arbitrarily assigned average contact energy.

\subsection{Generation of Datasets}

For each coefficient of friction $\mu$ and force
cutoff configuration a set of 1000 horizontally periodic lattices of
100 rows by 100 columns with an applied load of 1000 N (the unit is
arbitrarily applied for the benefit of the applied limits) is
generated and averaged over to yield the probability distributions of
vertical and horizontal force components.  The load is distributed on
the top layer via a normal distribution centered about the average
force of 10 N with a standard deviation of 5 N.  Normalization of both
the vertical and horizontal force components is performed relative to
the average vertical force (10 N) imposed on a site by the load on
the system.  The lattice size is chosen to minimize computational time
as the failure rate (number of lattices which failed to run to
completion divided by number of lattices started) in lattice
generation grew significantly with increased vertical size despite the
measures taken to increase overall yield described in
appendix~\ref{app:PFB}.  However, as convergence of the distributions
proves to be fairly rapid, larger lattices are unnecessary.

Data sets with static coefficients of friction $\mu = 0$,
$\mu=0.1$, and  $\mu=0.2$ under various cutoff schemes are used.
Configurations for $\mu=0$ and $\mu=0.2$ are also generated 
without an applied force cutoff in order to demonstrate the necessity
of cutoff implementation.  In setting the sharp force cutoff, upper
limits of 50 N and 100 N are set on the normal forces.  Angle
cutoffs are implemented for a range about $60^{\circ}$ to simulate a
triangular packing.  For the soft cutoff, eq. (\ref{eq:soft}), we use
the values based on the physical properties of soda lime-silica float
glass ($\sigma = 0.23$, $E = 7.2 \times 10^{10}$ Pa) assuming a radius
$R$ of 1.75 mm, yielding $A(\sigma,E,R) = 1.5 \times 10^{-7} \mathrm{J
\cdot N}^{-\frac{5}{3}}$ for eq. (\ref{eq:contact_U}).
Total input energies of 1 and 5 J are considered, leading to  $U_{0} =
5 \times 10^{-5}$ J and  $U_{0} = 2.5 \times 10^{-4}$ J, respectively.

\section{Results}
\label{sec:Results}
We find the probability distribution $P(v)$ of normalized vertical
forces $v$ shows remarkable robustness and appears to be independent of the
coefficient of friction $\mu$ and of choices of large force cutoff.  In
contrast, the probability distribution $P(h)$ of normalized horizontal
forces $h$ exhibits changes in functional form with variation in both
$\mu$ and with cutoff choice.  Horizontal forces, in general, are of
larger magnitude than vertical forces.

\begin{figure}
\centering
\mbox{
\subfigure[]{\epsfig{file=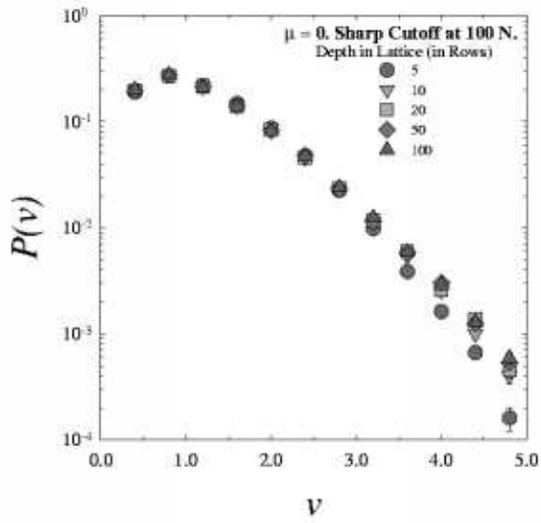, width=3.375in}}
\subfigure[]{\epsfig{file=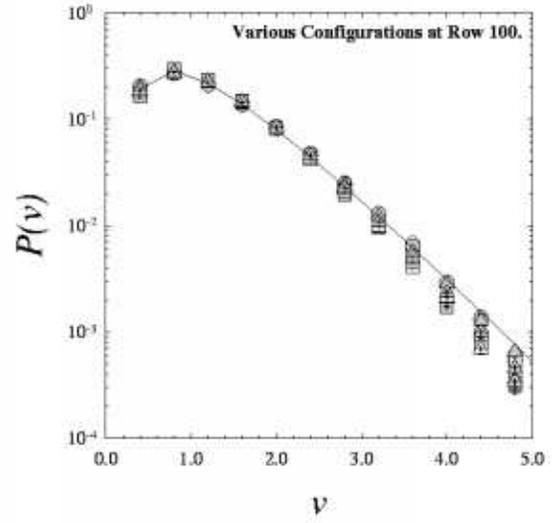, width=3.375in}}
}
\caption{(a) Probability distribution $P(v)$ of normalized vertical
force $v$ at increasing depths of a non-frictional $(\mu = 0)$
lattice.  Force limit has been arbitrarily set to 100 N (unnormalized)
for normal forces.  Convergence occurs fairly rapidly (within $\sim
10$ layers).  Similar results are seen for other values of $\mu$ and
force cutoff configurations.  (b) Probability distributions $P(v)$ for
normalized vertical force $v$ at depth 100 for various values of $\mu$
and force cutoff configurations.  The symbols are defined in table
\ref{tbl:symbol}.  Functional form of the distributions is invariant
with respect to configuration.  The force distributions are very
similar to that of the scalar q model with $N=2$, shown as the solid
line, which is appropriate for this geometry~\cite{Coppersmith:q:1996}.}
\label{fig:vertical}
\end{figure}

\begin{figure}
\centering
\epsfig{file=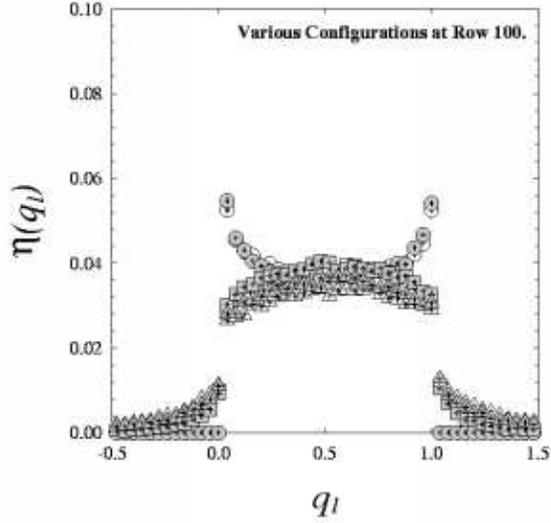, width=3.375in}
\caption{Histogram of $q_l$ values describing
the redistribution of the vertical component of force for various
values of $\mu$ and force cutoff configurations at a depth of 100 rows,
denoting the fraction of weight supported by the leftward neighbor of
a site.  Symbols used are the same as for figure \ref{fig:vertical}(b)
(defined in table \ref{tbl:symbol}) except only the larger, less
restrictive cutoff values are used (Sharp Cutoff at 100 N and Soft
Cutoff with 5 J).  Bin size, $\Delta q$, is 0.04 and the result is
placed at the right edge of the bin.  The distribution is nearly
uniform except for a slight increase in probability near values of 0
and 1 as compared to 0.5. for non-frictional configurations and a
slight decrease when friction is present.  Because of the admission of
friction in the vector model, $q$ values outside of the range $[0,1]$
are possible as frictional forces may cause a contact to have an
upward net force.  Roughly 5-20\% of $q$'s in $\mu \neq 0$
configurations are found to lie outside the range $[0,1]$ with a
smaller number found for more restrictive force cutoff configurations
and a larger number for less restrictive cutoffs.  Non-frictional
$(\mu = 0)$ configurations all have $q$ values lying in the range
$[0,1]$.}
\label{fig:q}
\end{figure}

Convergence of the probability distributions of normalized vertical
force component $v$ is fairly rapid, on the order of 10 rows.  $P(v)$
versus depth in lattice is shown in figure \ref{fig:vertical}(a)
for a non-frictional, sharp cutoff limit configuration.  Similar
results for $P(v)$ are seen for all generated configurations with little
variation in functional form; the distributions at row 100 are shown
in figure \ref{fig:vertical}(b) with the symbol key for the various
configurations shown in table \ref{tbl:symbol}.  An exponential tail
for $P(v)$ is seen for larger values of $v$ and a ``dip'' in $P(v)$ is
seen for small values of $v$.  The observed $P(v)$ is very similar to
that obtained from the scalar $q$ model with uniform $q$ distribution
for an $N=2$ (two-dimensional) system~\cite{Coppersmith:q:1996}
\begin{equation}
P(v) = 4 v^2 e^{-2v}.
\end{equation}
\noindent The distributions of $q$ values describing the redistribution
of the vertical forces, shown in figure \ref{fig:q}, are nearly uniform
with a slight increase in probability near values of 0 and 1 as
compared to 0.5 for $\mu = 0$ and a decrease when $\mu > 0$. In
addition, configurations with $\mu > 0$ in the vector model do  allow
for $q$ values outside the range $[0,1]$ due to mostly horizontal
normal forces with a corresponding $\eta < 0$.  The net vertical
component of the input force at a site is still kept positive.  For
the $\mu \neq 0$ configurations examined, roughly 5-20\% of the $q$
values are found to lie outside the range $[0,1]$ with a smaller
fraction found in more restrictive force cutoff configurations and a
larger fraction for less restrictive cutoffs.  Non-frictional $(\mu =
0)$ configurations all have $q$ values lying in the range $[0,1]$.  

\begin{figure}
\centering
\mbox{
\subfigure[]{
\epsfig{file=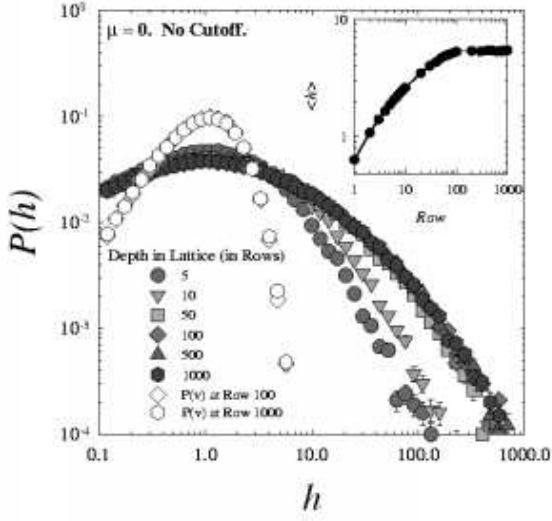,width=3.37in}
}
\subfigure[]{
\epsfig{file=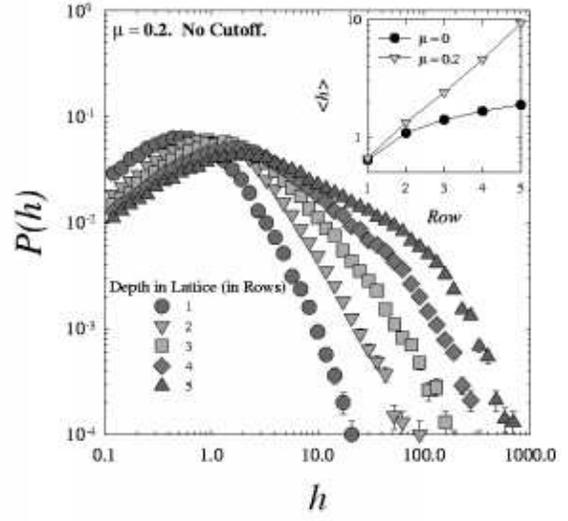,width=3.37in}
}
}
\caption{Results for lattices without imposed force cutoff.  We see
that the probability distribution $P(h)$ for normalized horizontal
forces $h$ broadens and shifts toward larger values of $h$ with depth
while $P(v)$ for normalized vertical forces $v$ remains unchanged.
The distribution $P(h)$ in the non-frictional $(\mu=0)$ case appears
to converge after about 100 layers, reaching upper limit of
$\left<h\right>/\left<v\right> \sim 5.5$, as shown in the inset.   The
distribution does not widen any further--- for large
$F^{in}_x/F^{in}_y$ ratios at individual sites, no valid solutions of
force and torque balance, eqs. (\ref{eq:balance}), under the
constraint of non-tensile forces, eq. (\ref{eq:non-tensile}), exist.
Although this appears to be an intrinsic limit, it results in
physically unreasonable values of force and is far less
effective when $\mu > 0$.  Results shown in (b) demonstrates that the
addition of friction greatly enhances the anisotropy.  Lattice
generation with friction present was limited to 5 rows due to an
increased rate of failure.  Deeper lattices are possible; however, the
number of points sampled in configuration space increases
significantly.}
\label{fig:nightmare}
\end{figure}

\begin{figure}
\centering
\mbox{
\subfigure[]{\epsfig{file=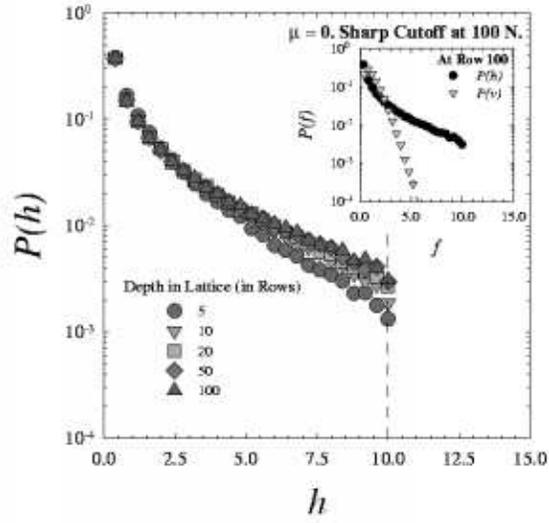, width=3.375in}}
\subfigure[]{\epsfig{file=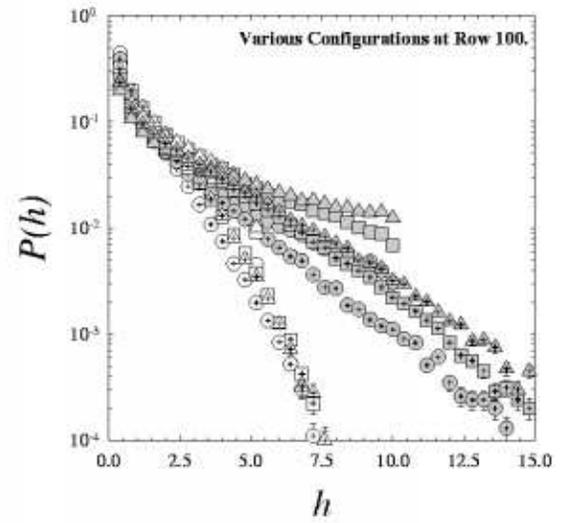, width=3.375in}}
}
\caption{(a) Probability distribution $P(h)$ of normalized horizontal
force $h$ at increasing depths for a non-frictional $(\mu = 0)$
lattice.  The normalization factor used is the same as for the
vertical component of force.  Force limit has been arbitrarily set to
100 N (unnormalized) for normal forces and the distribution terminates
abruptly near $h=10$ (100 N).  The distribution has not
converged and appears to be widening, indicating the average
horizontal force is increasing with depth. The functional form is
unlike the vertical and is on a larger scale as can be seen in the
inset.  (b) Probability distribution $P(h)$ at depth 100 for various
values of $\mu$ and force cutoff configurations.  Symbols used are the
same as for figure \ref{fig:vertical}(b) (defined in 
table \ref{tbl:symbol}).  Functional form varies with the imposed
limits.  While $P(h)$ curves for the sharp cutoff terminates at $h=5$
and  $h=10$, the soft cutoff Boltzmann-like limits do not.}  
\label{fig:horizontal}
\end{figure}

\begin{figure}
\mbox{
\subfigure[q Model]{\epsfig{file=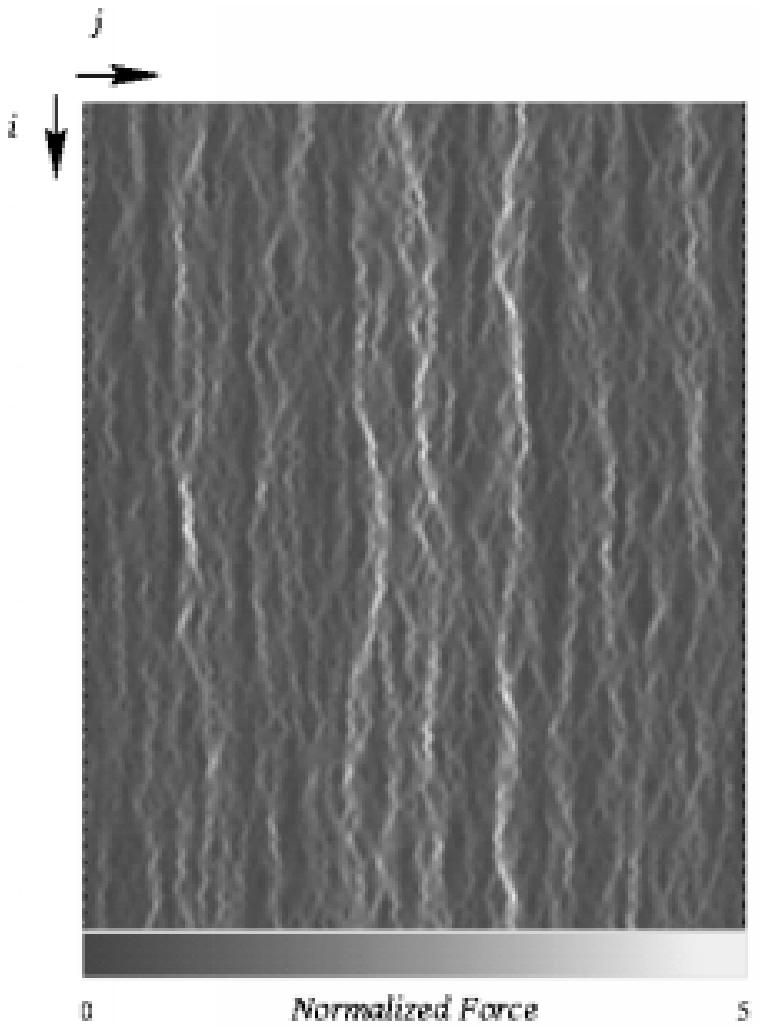, width=2.2in}}
\subfigure[Vector Model: Vertical Force Component]{
\epsfig{file=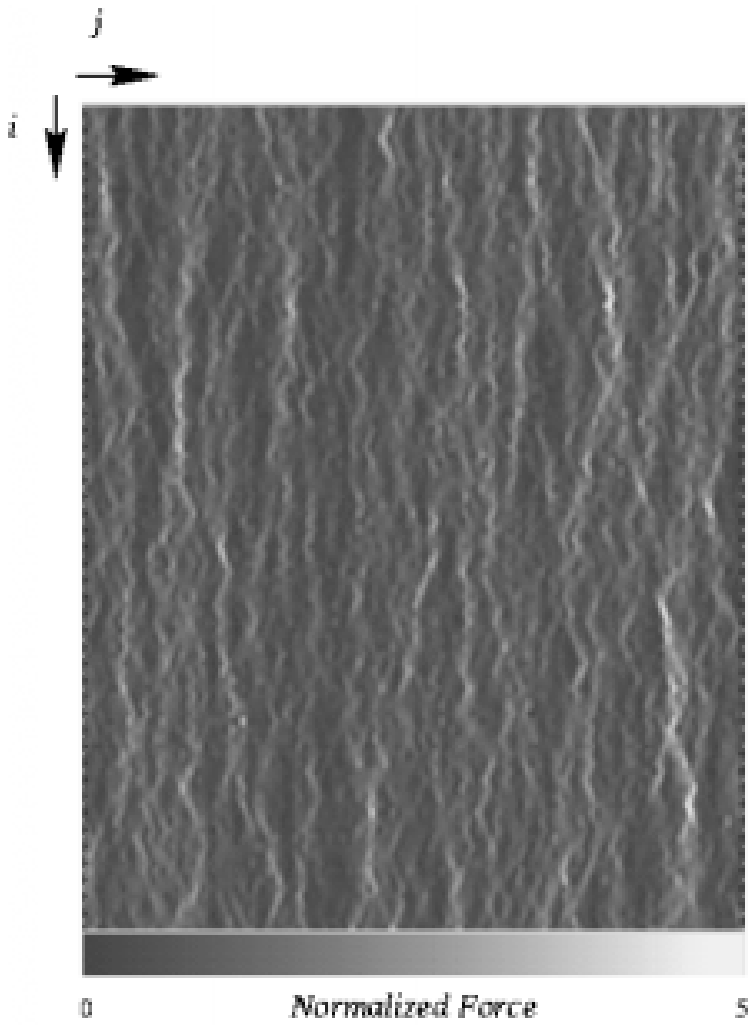, width=2.2in}}
\subfigure[Vector Model: Horizontal Force Component]{
\epsfig{file=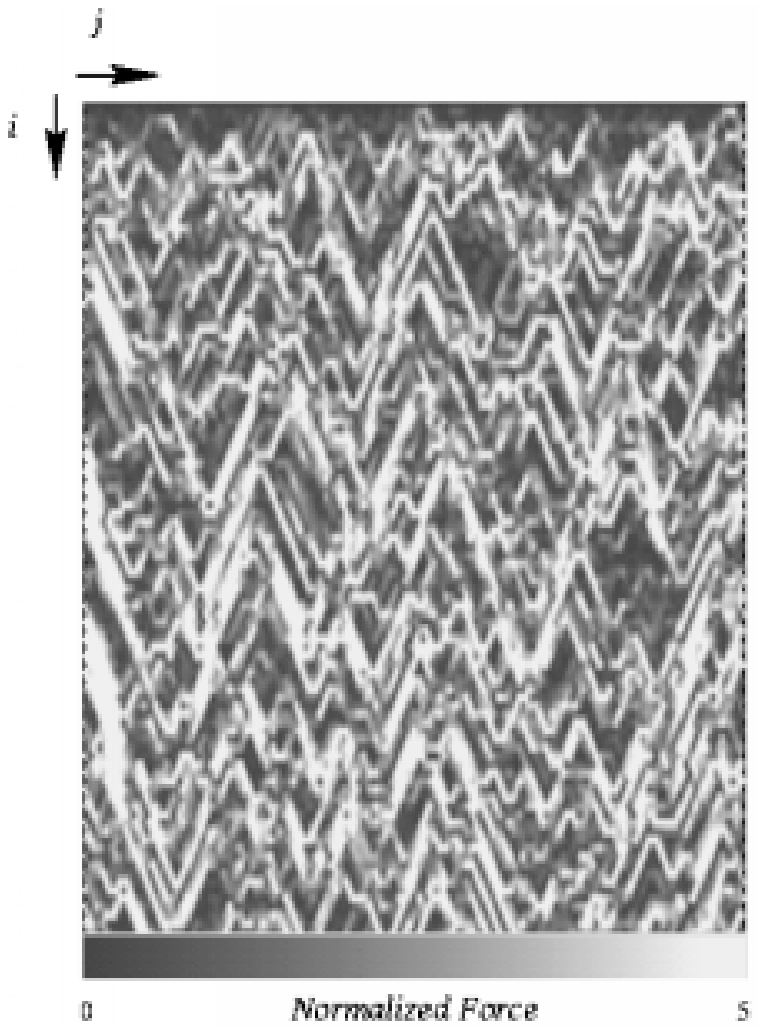, width=2.2in}}
}
\caption{Grayscale plot of representative 100 by 100 lattices for
the q model with uniform q distribution and the vector model with
$\mu=0.2$ and a sharp cutoff of 100 N for normal forces.  A load of
1000 N has been applied at the top of the lattices.  The darkness or
brightness of a site corresponds to the magnitude of the normalized
force component.  Normalized force components with magnitudes greater
than 5 have been clipped at 5.  Both the vertical and horizontal force
components are normalized by the same value.  The qualitative
difference between the two components for the vector model are readily
apparent; the horizontal fluctuations are of a much larger scale and
form ``V''-shaped chains reminiscent of light cones while the vertical
fluctuations resemble those found in the q model.} 
\label{fig:sample.lattice}
\end{figure}

In contrast to the vertical force distribution $P(v)$, the
probability distribution $P(h)$ of normalized horizontal force 
component $h$ shows great variation in form with changes in the
imposed force limit.  We first examine the $P(h)$ distributions 
for lattice sets with no imposed force cutoff and $\mu=0$ and $\mu=0.2$,
shown in figure \ref{fig:nightmare}.  We see the distributions
spreading toward larger values of $h$ as we progress deeper into the
lattice while $P(v)$ remains bounded and robust.  When $\mu = 0$,
$P(h)$ converges, but at values of the horizontal force much greater
than the vertical force, as seen in \ref{fig:nightmare}(a).  We find
that this convergence is mainly to a decrease in the number of valid 
solutions of force and torque balance, eqs. (\ref{eq:balance}), that
satisfy the non-tensile and frictional constraints given by
eqs. (\ref{eq:constraints}) as the ratio of input force components
$F^{in}_x / F^{in}_y$ grows larger.  Although this process appears to
provide an intrinsic limit on forces, it results in physically
unreasonable values of force and fails altogether to limit the magnitude
of horizontal forces when $\mu > 0$.  The addition of friction greatly
enhances the anisotropy--- the generation of large horizontal forces
in configurations without friction only occurs at nearly horizontal
values of $\varphi_{l,r}$, while the addition of friction allows for
the generation to occur for a larger region of angles.  More
precisely, the $\Psi = 0$ boundary for $\mu = 0$ occurs only at
\begin{equation}
\varphi_l + \varphi_r = 0,
\end{equation}
while it varies for $\mu = 0.2$ according to
\begin{equation}
0 \leq \varphi_l + \varphi_r \leq 22.6^\circ.
\end{equation}
The rapid growth in the magnitude of $h$ values prevents the
generation of large frictional lattices without imposed force cutoffs
to take place in a reasonable amount of time.  When a sharp cutoff in
the normal force is imposed, the probability distribution P(h), as
seen in figure \ref{fig:horizontal}(a) cuts off abruptly, which is
unlikely to be realized in a physical packing.  The use of angle
cutoffs was unsatisfactory as lattice generation for this scheme
was consistently terminated due to angles being driven to the cutoff
values due to our choice of parameterization and random value
selection.  Configurations obtained using the soft cutoff from
eq. (\ref{eq:soft}), shown in figure \ref{fig:horizontal}(b), 
exhibit differing functional forms for the probability distribution as
the cutoff is changed with  broadening occurring at the tail with
increased input energy.  For small values of $h$, the distribution
appears to be unaffected by changes in the choice of cutoff.
Increasing $\mu$ serves to broaden the distribution $P(h)$.  As
force components are normalized by the same factor, it can be readily
seen that the scale of horizontal forces is larger than the vertical
as shown in the inset of \ref{fig:horizontal}(a).
A grayscale plot of the forces on a representative sample lattice 
with $\mu = 0.2$ and a sharp cutoff at 100 N is shown in figure
\ref{fig:sample.lattice}.  It is obvious that the vertical and
horizontal force fluctuations differ qualitatively. 

Thus, although the imposition of force limits results in horizontal
force distributions which do not diverge as the depth is increased,
these limits serve only to mask the divergent behavior of the model:
the distribution of forces expands to fill the space allowed and the
distribution of horizontal forces exhibits a strong dependence on the
choice of cutoff scheme and value.

\section{Discussion}
\label{sec:Discussion}
The layer-by-layer vector model that we have investigated to model
forces in granular packings yields vertical and horizontal force
probability distributions which are of different functional forms and
scales, in contrast with the recent measurements by Mueth \textit{et
al.}~\cite{Mueth:1998}.  Moreover, while the distribution of vertical
forces is robust, the distribution of horizontal forces depends on the
details of our implementation. 

The asymmetry seen in the force distributions is a reflection of
the vertical-horizontal asymmetry inherent in any uni-directional
layer-by-layer model.  At the level of individual elements, we see
that large-magnitude horizontal forces are generated.  As these
forces accumulate, the resulting probability distribution of
horizontal force components is naturally skewed toward larger values. 
In contrast, large vertical forces are not generated.  The imposition
of limits on the vertical forces by the applied load and the
non-tensile force constraint is sufficient to limit the magnitude of
that force component.  The horizontal component has no 
similar constraints except those arbitrarily imposed by our cutoff
schemes.

This behavior is apparent in other implementations of layer-by-layer
models despite differing choices in parameterization and
configuration.  Eloy and Cl\'{e}ment~\cite{Eloy:1997} model the
packing by assuming a mono-disperse 2D array of hard cylinders
arranged in a triangular lattice.  The angle of contact between
cylinders in neighboring layers is fixed at slightly less than
$60^{\circ}$ measured with respect to the horizontal.  The
redistribution of forces at a cylinder is parameterized by the
coefficient of friction $\mu$ and the difference in
value of the horizontal force components transfered to the neighboring
sites in the row below, $p$.  They clearly note that for certain values of
$\mu$, valid values of $p$ are unreasonably large in magnitude and an
arbitrary cutoff, restricting the parameter space to be far away from
this divergent region, is imposed.  Socolar's $\alpha$
model~\cite{Socolar:1998} represents the packing with a lattice of
square cells with net normal forces, couples (torques), and tangential
forces represented on each edge. The redistribution of forces at a
cell is parameterized by  a triplet of values $(\alpha_0, \alpha_1,
\alpha_2)$ which are used to couple the torque and force balance.  The
distribution of net forces at every site is restricted to be in a
downward cone opening at $45^{\circ}$ by considering only frictional
forces acting in conjunction (the same direction vertically) with
normal forces.  This restriction guarantees successful lattice
generation but renders the force fluctuations anisotropic by
construction--- the magnitude of the horizontal force components is
artificially limited to less than that of the vertical.  We believe that
widening the cone by including frictional forces in opposition to normal 
forces would lead to the divergent behavior of horizontal force components.

\begin{figure}
\centering
\epsfig{file=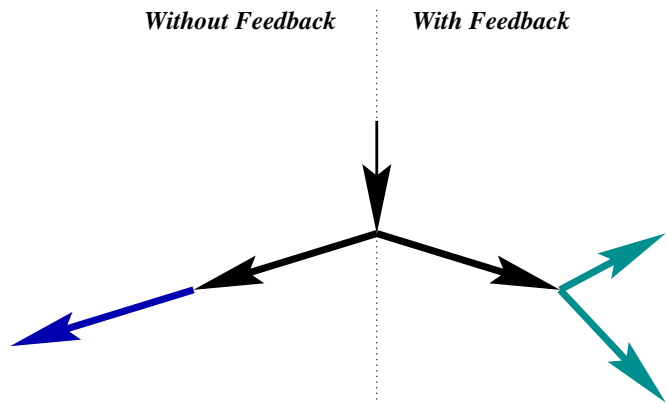, width=3.5in}
\caption{Schematic of feedback arising from a large horizontal force.
The left-hand side shows the force propagation as implemented in
layer-by-layer models.  The right-hand side demonstrates how the force
could be broken into upward- and downward-traveling components.}
\label{fig:feedback}
\end{figure}

Transmission of information regarding the production of
large-magnitude forces upwards into the already processed portion of
the lattice is a mechanism missing in layer-by-layer models which
could serve as means of suppressing large magnitude forces.  To
illustrate this point, figure~\ref{fig:feedback} shows the breakup of
a large horizontal force into upward- and downward-traveling
components .  In addition to reducing the magnitude of a single force
which would have otherwise persisted, this process yields an upward
force component that may instigate a rearrangement which then causes a
new redistribution of forces in the layers above and allow the site
generating the force to ``relax'' to a less stressful configuration or
generate a loop which isolates the force chain.  One must consider
some sort of feedback mechanism to enable corrective rearrangements
and redistribution of forces to be made.

Implementation of a feedback mechanism within the context of a
layer-by-layer method is not a trivial matter.  While a
pseudo-feedback mechanism (see appendix \ref{app:PFB}) is used in the
generation of the sets of lattices analyzed here and serves the purpose
of increasing the likelihood of finding an allowable configuration, it
still maintains the downward propagation of actual force information.  

In summary, the intrinsic asymmetry of layer-by-layer vector models
results in asymmetries in force distributions that are not seen in
experiment.  This failure of the model indicates that constructing a
statistical characterization of vector force transmission in granular
media requires a better understanding of the vectorial nature of force
redistribution, taking into account explicitly the symmetry properties
of the medium.

\acknowledgements
This work was supported primarily by the MRSEC program of the National
Science Foundation under Award Number DMR 9400379.  We are grateful to
D. Mueth, S. Nagel, H. Jaeger, R. Clelland, F. Radjai, and J. Socolar
for useful conversations.

\appendix
\section{Pseudo-Feedback and Increased Lattice Yield}
\label{app:PFB}

The generation of a sufficiently large set of lattices is essential in
performing any reasonable statistical analysis of the force
distribution in the model.  In the $q$ model, valid redistribution of
force at a site is guaranteed because the constraints on the choices
of $q$'s are independent of the force acting at a given site.  In the
vector model proposed here, as well as in \cite{Eloy:1997} and 
\cite{Socolar:1998}, the set of random statistical variables allowed 
at a site is dependent on the force acting upon it, as can be seen
by eqs. (\ref{eq:forces}) and (\ref{eq:constraints}) for
calculating the output normal forces and checking the non-tensile
constraint, respectively.  We choose to select the set of random
statistical variables for a site, test to see if the 
constraints are met, and discard and reselect a new set of
variables if the constraints are not satisfied.  Although seemingly
inefficient, this method is far simpler than encapsulating the
non-tensile and friction constraints within the choice of random
variables.  The lack of a guaranteed redistribution configuration is
the root cause of the failure in lattice generation in the vector
models we examined. 

We may reduce the rate of incidence of failure if we can
ensure that the inputs at the sites in the layer below that currently
being redistributed will be able to support valid force
redistributions themselves.  We do this by incorporating a simple
check in the selection process for the statistical variables 
$\varphi_{l,r}$ and $\eta_{l,r}$ for neighboring sites in the same layer.
If a valid redistribution is not possible, then all sites on the row are
subjected to a new redistribution.  We may extend this process to an
arbitrary number of resulting layers although this may significantly
increase the memory requirements for computation in addition to
complicating the layer-by-layer algorithm.  The essence of this
requirement is to capture the flavor of feedback and rearrangement,
albeit in an imprecise manner.  However, we have no guarantees
that a valid configuration will be possible. 

This pseudo-feedback (PFB) addition was set to a depth of two layers in the
process of generating the lattices used in this paper.  Typically,
most difficulties in redistribution may be resolved with a one-layer PFB
as this provides an immediate check for input forces leading to a
valid redistribution.  However, these valid redistributions may lead
to an invalid input force on the next layer down, a less likely but
still significant cause of failure.  The implementation of a
second layer for the PFB resolves this issue and allows for the
vast majority of lattices for our configurations of friction and force 
cutoff to run to completion.  Using PFB beyond two layers increases the
yield but not significantly.  

\section{Considerations Relevant for Symmetric Lattices: Four-site Lattices}
\label{app:Four}
We have examined exactly solvable lattice configurations to study the
interaction between force-balanced sites to gain insight into how
large-magnitude forces may persist in the lattice and what methods
beyond externally imposed force limits exist to prevent or remove
them.  Although they do not offer any statistical information, these
configurations offer further insight into the nature of the generation
and propagation of large magnitude horizontal forces by allowing
greater control over the parameters governing the forces in a packing
than is found in a stochastic layer-by-layer approach.

\begin{figure}
\centering
\epsfig{file=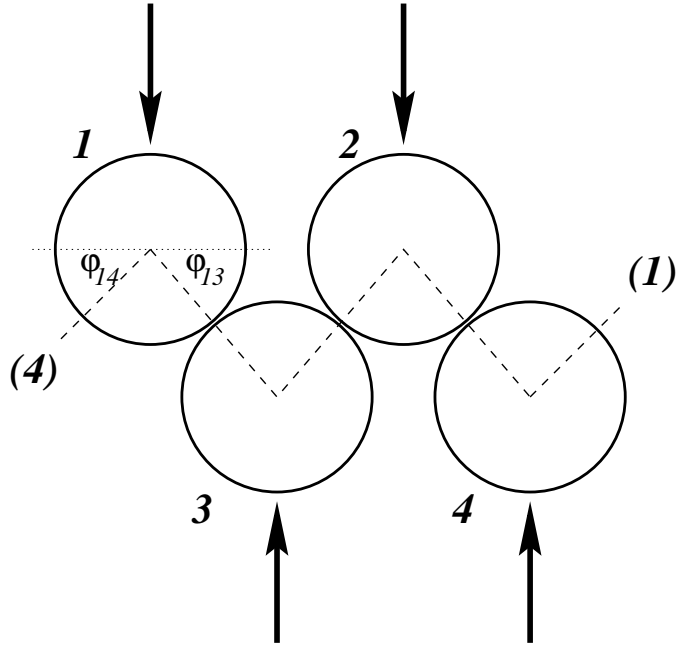, width=3.5in}
\caption{Schematic of the four-site triangular lattice.  External
vertical forces have been applied to each site.  The configuration is
made periodic by connecting sites 1 and 4 as shown.  Indexing of the
contact angles and effective friction coefficients refer to the
sites in contact (ie. $\varphi_{13}$ is the contact angle between
sites 1 and 3).  As angles are measured with respect to the horizontal,
$\varphi_{ij} = \varphi_{ji}$.}
\label{fig:4site.triangular}
\end{figure}

We find that four-site lattices represent the smallest meaningful unit of
study.  The two configurations which maintain the coordination
number four are the two-layer triangular lattice and the quadrilateral
shown in figures \ref{fig:4site.triangular} and \ref{fig:4site.quad},
respectively.  We examine both horizontally periodic and fixed-wall cases. 

\subsection{Two-Layer Triangular Lattice}

Placing the sites in the familiar triangular lattice shown in figure
\ref{fig:4site.triangular}, vertical forces are applied to each site.
Because of the symmetry of the system, two angles and a single
internal friction coefficient are sufficient to specify the
redistribution of forces for the internal contacts.  Assuming that the
known angles are $\varphi_{13}$ and $\varphi_{23}$ and the known
friction coefficient is given by $\eta \equiv \eta_{13}$, the normal
forces for the periodic case are 
\begin{mathletters}
\begin{eqnarray}
F_{13} & = & F_3^{in} \cos \varphi_{23}
         \left(\sin (\varphi_{13} + \varphi_{23}) +
         \mu \eta
         \left(1 + \cos (\varphi_{13} + \varphi_{23}) \right)
         \right)^{-1},
\\
F_{23} & = & F_{13} \left(\cos \varphi_{13} -
                  \mu \eta \left(\sin \varphi_{13} - \sin \varphi_{23}\right)
                \right) / \cos \varphi_{23},
\\
F_{24} & = & F_{13} \left(\cos \varphi_{13} -
                  \mu \eta \left(\sin \varphi_{13} - \sin \varphi_{24}\right)
                \right) / \cos \varphi_{24},
\\
F_{14} & = & F_{13} \left(\cos \varphi_{13} -
                  \mu \eta \left(\sin \varphi_{13} - \sin \varphi_{14}\right)
                \right) / \cos \varphi_{14},
\end{eqnarray}
\end{mathletters}
where $\varphi_{14}$ and $\varphi_{24}$ are defined implicitly by
\begin{mathletters}
\begin{eqnarray}
\frac{F_3^{in}}{F_2^{in}} & = &
\frac{\cos \varphi_{24}
      \left(\sin (\varphi_{13} + \varphi_{23}) +
            \mu \eta \left(1 + \cos (\varphi_{13} + \varphi_{23}) \right)
      \right)}
     {\cos \varphi_{13} \sin (\varphi_{23} + \varphi_{24}) +
                   \mu \eta
          \left(\cos \varphi_{23} \left(1 + \cos (\varphi_{23} + \varphi_{24})
                                        \right)
                         - \frac{\sin (\varphi_{23} + \varphi_{24})}
                                {\sin \varphi_{13} - \sin \varphi_{23}}
                   \right)},
\\
\frac{F_1^{in}}{F_3^{in}} & = &
\frac{\cos \varphi_{23} \left(\sin (\varphi_{13} + \varphi_{14}) +
                           \mu \eta \left(1 + \cos (\varphi_{13} + \varphi_{14})
                                    \right)
                     \right)}
     {\cos \varphi_{14} \left(\sin (\varphi_{13} + \varphi_{23}) +
                           \mu \eta \left(1 + \cos (\varphi_{13} + \varphi_{23})
                                    \right)
                     \right)}.
\end{eqnarray}
\end{mathletters}

Assuming uniform vertical inputs, the periodic case admits solutions
near the single-site $\Psi=0$ boundary with the balancing of external
forces remaining perfectly satisfied while the magnitudes of the
internal normal and frictional forces grow unbounded.  A similar
result exists in the non-uniform case once external force and torque
balance is taken into account.

The fixed horizontal boundaries case is derived by disconnecting sites
1 and 4 and fixing the direction of the normal forces to be normal to
the wall.  As the magnitude of the sum of the horizontal components of
each normal and frictional force for each contact must be equal, any
restriction placed on the at the wall (ie. loading) will place limits
on the internal redistribution.  By removing the asymmetry of loading
in this configuration, we can prevent the formation of the large
magnitude forces.  However, we must be able to specify all the forces
along the boundary of the system.  If we are only given the vertical
loading, then any restrictions placed on the horizontal force
component will be arbitrary.

This result may indicate that successful implementation of
probabilistic vector models of force fluctuations requires better
understanding of the role of the boundaries.  This question is also 
crucial to understanding whether the equations underlying these force
distributions are elliptic or hyperbolic~\cite{Claudin:1998}.

\subsection{Quadrilateral Lattice}

\begin{figure}
\centering
\epsfig{file=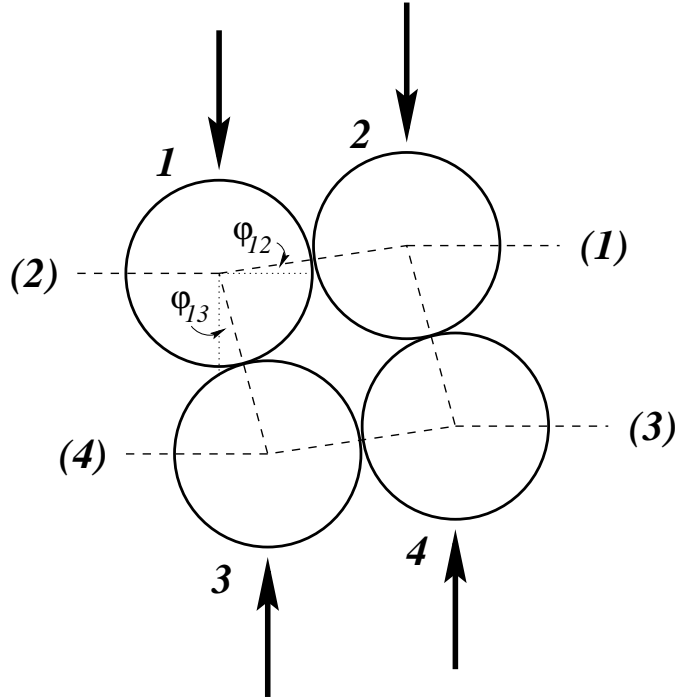, width=3.5in}
\caption{Schematic of the four-site quadrilateral lattice.  External
vertical forces have been applied to each site.  The configuration is
made periodic by connecting sites on the same vertical layer as shown
(site 1 to site 2, site 3 to site 4).  Contact angles are measured by
their deviation from a square (ie. $\varphi_{12}$ is the deviation
from the horizontal and $\varphi_{13}$ is the deviation from the
vertical and $\varphi_{ij} = -\varphi_{ji}$).}
\label{fig:4site.quad}
\end{figure}

The quadrilateral configuration shown in figure \ref{fig:4site.quad}
offers the advantage of having a ``rotational'' symmetry
lacking in both the single-site and triangular lattice configurations.
We have implemented an explicit loop to observe its effect on the
internal force redistributions and hopefully allow us insight into the
role of this symmetry in the system.  

We apply vertical external forces to each site.  Periodic boundary
conditions are created by pairing the horizontal inputs of sites on
the same vertical ``level'' so that they are equal in magnitude and
have friction coefficients acting on each site of the pair equally and
oppositely.  Although the equations of force balance for the system
are easily derived, valid configurations are more readily determined
numerically. Unbounded horizontal force solutions are supported--- the 
redistribution pair consists of the periodic horizontal and the 
internal force connecting same-level sites.  The internal loop
prevents these forces from being passed between vertical levels due to
the direction of the tangential frictional forces within the loop being
prevented from aligning adversely in neighboring loop sites.

The addition of fixed horizontal boundaries leads to results similar
to the triangular lattice configuration.  We see that the introduction
of internal symmetry to the system is not enough--- we are still subject 
to the effects of an externally imposed asymmetry.  However, we do gain
the ability to isolate the effects, in this case over a layer, as
opposed to its propagation throughout the system.

\clearpage

\clearpage

\clearpage

\clearpage

\clearpage

\clearpage

\clearpage

\clearpage

\begin{table}
\caption{Symbol key for figures \ref{fig:vertical}(b), \ref{fig:q},
and \ref{fig:horizontal}(b) which are plots consisting of various
configurations of coefficient of static friction $\mu$ values and
force cutoff schemes.}
\label{tbl:symbol}
\begin{tabular}{lcccc}
& 
\multicolumn{2}{c}{Sharp Cutoff} & 
\multicolumn{2}{c}{Soft Cutoff} \\
& 50 N & 100 N & 1 J & 5 J \\
$\mu = 0$ & 
\epsfig{file=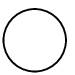, width=0.125in} &
\epsfig{file=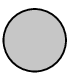, width=0.125in} &
\epsfig{file=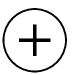, width=0.125in} &
\epsfig{file=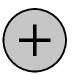, width=0.125in} \\
$\mu = 0.1$ & 
\epsfig{file=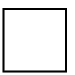, width=0.125in} &
\epsfig{file=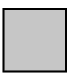, width=0.125in} &
\epsfig{file=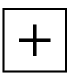, width=0.125in} &
\epsfig{file=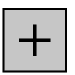, width=0.125in} \\
$\mu = 0.2$ & 
\epsfig{file=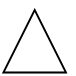, width=0.125in} &
\epsfig{file=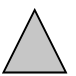, width=0.125in} &
\epsfig{file=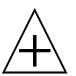, width=0.125in} &
\epsfig{file=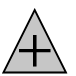, width=0.125in}
\end{tabular}
\end{table}


\begin{thebibliography}{10}

\bibitem{Liu:q:1995}
C. Liu {\it et~al.}, Science {\bf 269},  513  (1995).

\bibitem{Coppersmith:q:1996}
S.~N. Coppersmith {\it et~al.}, Phys. Rev. E {\bf 53},  4673  (1996).

\bibitem{Dantu:1967}
P. Dantu, Ann. Ponts Chauss. {\bf IV},  193  (1967).

\bibitem{Drescher:1972}
A. Drescher and G. de~Josselin~de Jong, Jnl. Mech. Phys. Solids {\bf 20},  337
  (1972).

\bibitem{Travers:1987}
T. Travers {\it et~al.}, Europhys. Lett. {\bf 4},  329  (1987).

\bibitem{Edwards:1996a}
S.~F. Edwards and C.~C. Mounfield, Physica A {\bf 226},  1  (1996).

\bibitem{Mounfield:1996}
C.~C. Mounfield and S.~F. Edwards, Physica A {\bf 226},  12  (1996).

\bibitem{Edwards:1996b}
S.~F. Edwards and C.~C. Mounfield, Physica A {\bf 226},  25  (1996).

\bibitem{Wittmer:1996}
J.~P. Wittmer, P. Claudin, M.~E. Cates, and J.-P. Bouchaud, Nature {\bf 382},
  336  (1996).

\bibitem{Wittmer:1997}
J.~P. Wittmer, M.~E. Cates, and P. Claudin, J. Phys. I France {\bf 7},  39
  (1997).

\bibitem{Claudin:1998}
P. Claudin, J.-P. Bouchaud, M.~E. Cates, and J.~P. Wittmer, Phys. Rev. E {\bf
  57},  4441  (1998).

\bibitem{Hemmingsson:1996}
J. Hemmingsson, Physica A {\bf 230},  329  (1996).

\bibitem{Hemmingsson:1997}
J. Hemmingsson, H.~J. Herrmann, and S. Roux, J. Phys. I France {\bf 7},  291
  (1997).

\bibitem{Ammi:1987b}
M. Ammi, D. Bideau, and J.~P. Troadec, J. Phys. D {\bf 20},  424  (1987).

\bibitem{Baxter:1997}
G.~W. Baxter,  in {\em Powders and Grains 97}, edited by R.~P. Behringer and
  J.~T. Jenkins (Balkena, Rotterdam, 1997), pp.\ 345--348.

\bibitem{Dantu:1957}
P. Dantu,  in {\em Proceedings of the 4th International Conference on Soil
  Mechanics and Foundation Engineering}, London, 1957 (Butterworths, London,
  1958), pp.\ 144--148.

\bibitem{Wakabayashi:1959}
T. Wakabayashi,  in {\em Proceedings of the 9th Japan National Congress for
  Applied Mechanics}, Tokyo, 1959 (Science Council of Japan, Tokyo, 1960), pp.\
  133--140.

\bibitem{Travers:1988}
T. Travers {\it et~al.}, J. Phys. (France) {\bf 49},  939  (1988).

\bibitem{Howell:1997}
D. Howell and R.~P. Behringer,  in {\em Powders and Grains 97}, edited by R.~P.
  Behringer and J.~T. Jenkins (Balkena, Rotterdam, 1997), pp.\ 337--340.
See Ref.\ \cite{Baxter:1997}.

\bibitem{Mueth:1998}
D.~M. Mueth, H.~M. Jaeger, and S.~R. Nagel, Phys. Rev. E {\bf 57},  3164
  (1998).

\bibitem{Miller:1996}
B. Miller, C. O'Hern, and R.~P. Behringer, Phys. Rev. Lett. {\bf 77},  3110
  (1996).

\bibitem{Radjai:1996}
F. Radjai, M. Jean, J.~J. Moreau, and S. Roux, Phys. Rev. Lett. {\bf 77},  274
  (1996).

\bibitem{Thornton:1998}
C. Thornton and S.~J. Antony, submitted to Phil Trans Roy Soc A (1998).

\bibitem{Clelland:unpub}
R. Clelland (unpublished).

\bibitem{Claudin:1997}
P. Claudin and J.-P. Bouchaud, Phys. Rev. Lett. {\bf 78},  231  (1997).

\bibitem{Nicodemi:1998}
M. Nicodemi, Phys. Rev. Lett. {\bf 80},  1340  (1998).

\bibitem{Eloy:1997}
C. Eloy and C. Cl\'{e}ment, J. Phys. I France {\bf 7},  1541  (1997).

\bibitem{Socolar:1998}
J.~E.~S. Socolar, Phys. Rev. E {\bf 57},  3204  (1998).

\bibitem{Pitman:1998}
E.~B. Pitman, Phys. Rev. E {\bf 57},  3170  (1998).

\bibitem{Kenkre:1998a}
V.~M. Kenkre, J.~E. Scott, E.~A. Pease, and A.~J. Hurd, Phys. Rev. E {\bf 57},
  5841  (1998).

\bibitem{Scott:1998b}
J.~E. Scott, V.~M. Kenkre, and A.~J. Hurd, Phys. Rev. E {\bf 57},  5850
  (1998).

\bibitem{Landau:Elastic}
L.~D. Landau and E.~M. Lifshitz, {\em Theory of Elasticity}, Vol.~7 of {\em
  Course of Theoretical Physics}, 3rd  ed. (Butterworth-Heinemann, Oxford,
  1986).

\end{thebibliography}
\end{document}